\documentclass[a4paper,11pt]{article}
\pdfoutput=1 % if your are submitting a pdflatex (i.e. if you have
\usepackage{jinstpub} % for details on the use of the package, please
\usepackage{mhchem}
\usepackage{subcaption}
\usepackage{multirow}
\title{Scintillation properties of \ce{(Zn_{0.9} Pb_{0.1})(W_{0.9} Mo_{0.1})O4} and \ce{(Zn_{0.9} Cd_{0.1})(W_{0.9} Mo_{0.1})O4} mixed crystals}

\author[a]{E.N.~Galashov,}
\author[a,b,1]{D.V.~Matvienko\note{Corresponding author},}
\author[a]{V.A.~Moskovskyh,}
\author[b]{B.I.~Sikach}
\author[a,b]{and B.A.~Shwartz}

\affiliation[a]{Novosibirsk State University, \\ 630090, Pirogova st. 2, Novosibirsk, Russia}
\affiliation[b]{Budker Institute of Nuclear Physics SB RAS,\\ 630090, Lavrentieva av., 11, Novosibirsk, Russia}

\emailAdd{d.v.matvienko@inp.nsk.su}

\abstract{Scintillation properties of \ce{(Zn_{0.9} Pb_{0.1})(W_{0.9} Mo_{0.1})O4} and \ce{(Zn_{0.9} Cd_{0.1})(W_{0.9} Mo_{0.1})O4} mixed crystals with doping of \ce{Eu, Sm, Pr, Ce, Sc, Yt} and \ce{Nb} are studied. Measurements of their light yields relative to pure \ce{ZnWO4} at room temperature, decay times and energy resolutions at 662 keV are presented. Emission spectra are obtained with ${}^{239}{\rm Pu}$ source of alpha particles.}

\keywords{Scintillators, scintillation and light emission processes (solid, gas and liquid scintillators); Interaction of radiation with matter}

\proceeding{Instrumentation of Colliding Beam Physics, February 24-28, 2020, Novosibirsk, Russia}

\begin{document}
\maketitle
\flushbottom

\section{Introduction}
\label{sec:intro}
A weak interacting massive particle (WIMP) is one of the proposed candidates for dark matter. Despite of the fact that most of the WIMP models are strongly constrained by direct, indirect and LHC searches, their parameter spaces are far from full exclusion. One of the direct search options are experiments with low temperature crystals, CRESST~\cite{cresst}, EDELWEISS~\cite{edelweiss}, CDMS~\cite{cdms} and proposed EURECA project~\cite{eureca}. The CRESST experiment operates with \ce{CaWO4} crystals as absorbers. In these crystals a WIMP particle could be elastically scattered on nuclei producing low energy nuclear recoils. The interaction produces heat and a small fraction of energy deposited is emitted as scintillation light. Since the light pulse produced differs for the different types of particle interaction ($\alpha/\beta$-particles, $\gamma$-quanta, neutrons), a powerful background discrimination with remarkable energy threshold and resolution could be achieved. The last results provided by the CRESST collaboration~\cite{cresst} show sensitivity for WIMP-nucleus cross section down to $10^{-6}$ pb. This value is a few orders of higher than one achieved by liquid-xenon experiments (LUX~\cite{lux}, XENON1T~\cite{xenon1t} and PandaX~\cite{pandax}). However cryogenic crystal experiments can be still competitive. The new EURECA project has been extensively discussed last years. It is a multi-target large experiment aimed to achieve a limit of spin-independent WIMP-nucleus cross section of $10^{-10}-10^{-11}$ pb. Such limit requires very low radioactive contamination of the materials used as well as high light yields at cryogenic temperatures. A family of heavy inorganic scintillators \ce{ABO4} (A=\ce{Zn, Ca, Cd} and B=\ce{Mo, W}) is considered as an attractive target~\cite{nagornaya}.

The other potential application of such crystals comes from experiments aimed to search for neutrinoless double beta decay ($0 \nu 2\beta)$. The great interest of such decay is explained by observed neutrino oscillations which prove nonzero neutrino masses. The $0\nu 2\beta$ decay can give a complementary information to neutrino oscillation experiments and establish the Majorana nature of neutrino ($\nu=\bar{\nu}$). In addition, an effective neutrino mass extracting from the $0\nu 2\beta$ decay can constrain the sterile neutrino parameter space~\cite{abada}. The AMoRE experiment~\cite{amore} aims to search for $0 \nu 2 \beta$ decay of \ce{ ^{100}Mo  } nuclei using \ce{ CaMO4 } scintillating crystals operating at cryogenic temperatures. The detection principle is similar to WIMP searches. The simultaneous measurement of light and phonon signals allows one to suppress a major alpha background occurring from radioactive contamination. 
The R\&D studies consider various other molybdate crystals for the experiment upgrade to avoid high-expensive depleting of \ce{ ^{40}Ca } isotopes, which produce an internal background.

A zinc tungstate (\ce{ ZnWO4 }) scintillator is also discussed as a very promising material for WIMP and $0\nu 2\beta$ searches.  
An advantage of these crystals is that their light yield for heavy particles (nuclear recoils, alpha) varies according to the orientation of the particle path into the crystal with respect to the crystallographic axes. Such tendency is absent for $e/\gamma$ radiation which is isotropic. In addition, the crystals show high radiopurity properties. It gives a possibility to search for diurnal modulation of WIMP direction with \ce{ ZnWO4 } crystals~\cite{cappella}.
A natural abundance of \ce{ ^{64}Zn } and \ce{ ^{186}W } isotopes is relatively large. The $2\nu 2\beta$ decay of \ce{ ^{186}W } is expected to be suppressed. It provides a suitable conditions to search for $0\nu 2\beta$ decays. The transition \ce{ ^{64}_{30}Zn -> ^{64}_{28}Ni } allows one to search for double electron capture and electron capture with positron emission. 

The promising properties of \ce{ ZnWO4 } crystals provide a strong motivation to further improve performance of this material by growth of mixed and co-doped crystals based on \ce{ ZnWO4 } wolframite crystal structure and \ce{CdMoO4} (or \ce{PbMoO4}) scheelite-type structure. 
In this study we consider the eight crystal compounds derived from \ce{ ZnWO4 } and \ce{ CdMoO4 } (or \ce{ PbMoO4 }) solid solutions and mixed with $90\%$ and $10\%$ proportions. The obtained mixed structures are doped with ($1$ at. \%) ion-activators. The molecular formula can be written as \ce{ (Zn_{0.9}Li_{0.01}X_{0.1})(W_{0.9}Mo_{0.1})O4 (Y)}, where \ce{ Y = Eu, Sm, Pr} for \ce{ X = Pb } (\ce { ZnWPbMoO4(Y) }) and \ce{ Y = Eu, Ce, Sc, Yt, Nb } for \ce{ X = Cd } (\ce{ ZnWCdMoO4(Y)}). In addition, one crystal without doping for \ce{ X = Cd } is used. Lithium ions are inserted to compensate a charge of ion-activators.
The \ce{ZnWPbMoO4(Y)} and \ce{ZnWCdMoO4(Y)} single crystals were grown in air on platinum crucible of $40$ mm in diameter according to the low-thermal-gradient Czochralski method in the Department of Applied Physics at Novosibirsk State University, Novosibirsk, Russia. The start growth temperature is $1182^{\circ}$C and $1192^{\circ}$C, which is lower than \ce{ZnWO4} ($1202^{\circ}$C). The growth direction is $[010]$. During the crystal growth the crystal-melt interface was polyhedral, represented mainly by plane $(010)$. 
A pure \ce{ZnWO4} crystal grown by the same technique is used as a reference for our study.

A comparative analysis of scintillation properties for pure \ce{ ZnWO4} and grown mixed samples is performed with the 662 keV photons from ${}^{137}{\rm Cs}$ source at room temperature. An emission spectrum is evaluated with ${}^{239}{\rm Pu}$ source of $\alpha$-particles. 

\section{Experimental details}
\label{sec:exp}
The \ce{ ZnWCdMoO4(Y) } crystals have cubic shapes with the size of $12$ mm while the \ce{ ZnWPbMoO4(Y) } crystals have quarter-cylinder shapes with the average cross section about $200$ mm${}^2$ and height $13$ mm.
The reference  \ce{ ZnWO4 } crystal has a cylindrical shape with the diameter of $30$ mm and height of $11$ mm. All samples are wrapped in two layers of 200 $\mu$m porous teflon coverage where the bottom face is kept to be opened. 

\begin{figure}[h]
    \centering
    \includegraphics[width=0.8\textwidth]{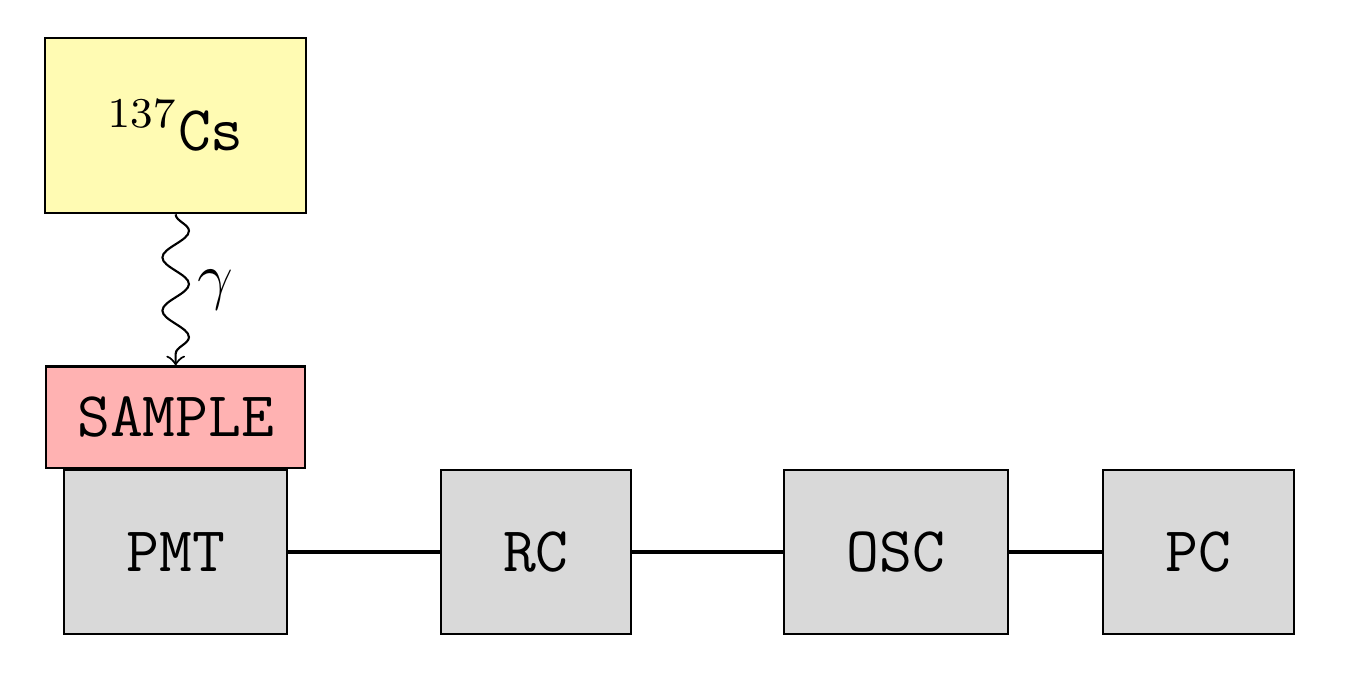}
    \caption{Principal scheme of experimental setup to measure the scintillation properties of the studied crystals.}
    \label{fig:setup}
\end{figure}

The measurements of scintillation characteristics are carried out using experimental setup shown in figure~\ref{fig:setup}.
The crystal is positioned on the photomultiplier tube (PMT) Hamamtsu R1847S with optical grease BC-630. The PMT has a bialkali photocathode and borosilicate glass window, linear focused ten-stage dynode system with a typical gain of the order of $10^7$ and spectral response in the range from $300$ nm to $650$ nm with maximum quantum efficiency of $28 \%$ at $420$ nm. The crystal is irradiated by a ${}^{137}{\rm Cs}$ source of $662$ keV photons and resulting scintillation light is detected.
The signal pulses are shaped by a RC-circuit with time constant of $750$ nsec to decrease the rate of false triggers and recorded by oscilloscope OWON TDS8204 within $150$ $\mu$s time gate. Pulse shape analysis includes calculation of baseline, determination of signal arrival, pulse height and integrated signal within tuned time window.

The PMT is calibrated by a single photoelectron method. An external light source CAEN LED driver SP5601 generating ultra-fast monochromatic pulses is used. The light intensity is adjusted in such a way that only one or several photoelectrons are emitted from the PMT photocathode.

\section{Decay time}
\label{sec:decay}
\begin{figure}[h]
    \centering
    \begin{subfigure}{0.5\textwidth}
    \centering
    \includegraphics[width=\textwidth]{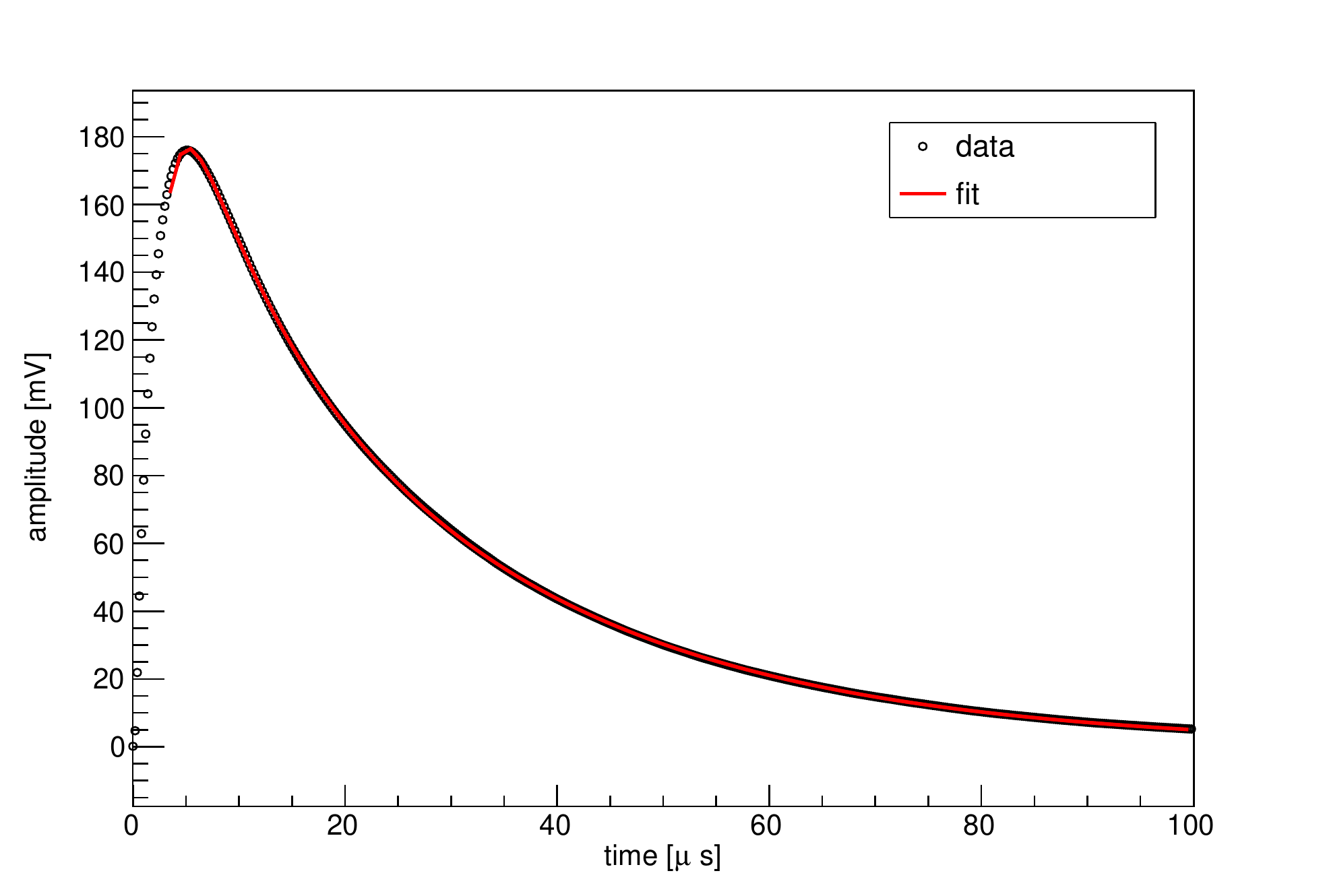}
    \caption{}
    \end{subfigure}%
    \begin{subfigure}{0.5\textwidth}
    \centering
    \includegraphics[width=\textwidth]{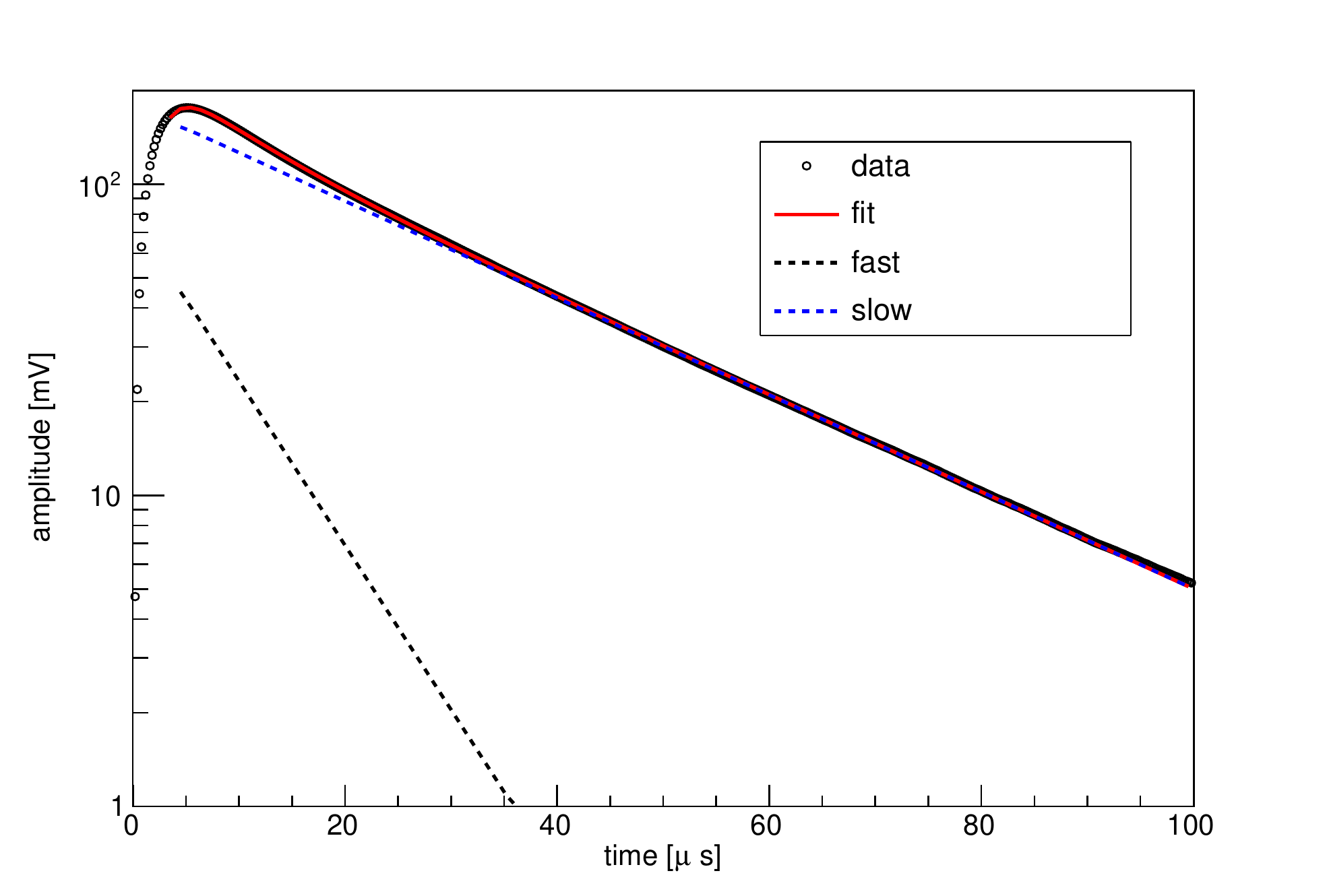}
    \caption{}
    \end{subfigure}
    \caption{Average pulse shape for the pure \ce{ZnWO4} crystal shown in (a) linear scale and (b) log scale. The red line shows the fit result, which gives $8.2$ $\mu$sec for the fast component (black dashed line) and $27.9$ $\mu$sec for the slow one (blue dashed line). Intensities of the components are $11\%$ and $89\%$, respectively.}
    \label{fig:decay}
\end{figure}

Decay time characteristics of the scintillation pulse of the studied crystals are obtained by the fit of the average pulse shape measured by oscilloscope when the crystal is irradiated by $662$ keV $\gamma$-quanta. The average pulse is built from individual pulses falling into the full absorption peak of the crystal energy spectrum. The pulse shape is fitted by two decay exponential functions with the lifetimes of the fast and slow components and one exponential function describing the rise of the pulse. The functions are corrected to the non-zero integration constant of electronics ($750$ nsec). The average pulse and result of the fit for the reference \ce{ZnWO4} crystal are shown in figure~\ref{fig:decay}.
The decay components and these intensities are in reasonable agreement with those obtained in ref.~\cite{danevich} (see table~\ref{tab:decayref}). Our measurement is not sensitive to the fast component of $0.7$ $\mu$sec in ref.~\cite{danevich} due to limited timing resolution in the pulse shape.
\begin{table}[h]
    \centering
    \caption{Decay times and their intensities of \ce{ZnWO4} scintillator for $662$ keV $\gamma$-quanta. The results are compared to those in ref.~\cite{danevich}.}
    \begin{tabular}{|l|l|l|}
    \hline
         Decay components & This work & Ref.~\cite{danevich}  \\ \hline
         Fast, $\mu$sec (\%) & --- & $0.7$ ($2$) \\
         Fast, $\mu$sec (\%) & $8.2$ ($11$) & $7.5$ ($9$) \\ 
         Slow, $\mu$sec (\%) & $27.9$ ($89$) & $25.9$ ($89$) \\ \hline
    \end{tabular}
    \label{tab:decayref}
\end{table}
The decay time constants of the \ce{ZnWPbMoO4} and \ce{ZnWCdMoO4} samples measured with the same method as for the pure \ce{ZnWO4} crystal are presented in table~\ref{tab:decay}. 
\begin{table}[h]
    \centering
    \caption{Decay times and their relative intensities (shown in percentage of the total intensity) of the \ce{ZnWPbMoO4} and \ce{ZnWCdMoO4} samples. The result for the reference \ce{ZnWO4} is also shown.}
    \begin{tabular}{|l|ll|ll||l|ll|ll|}
        \hline
        \multirow{2}{*}{Sample} & \multicolumn{2}{c|}{Fast} & \multicolumn{2}{c||}{Slow}
        & \multirow{2}{*}{Sample} & \multicolumn{2}{c|}{Fast} & \multicolumn{2}{c|}{Slow} \\
        & $\mu$sec & \% & $\mu$sec & \% & &  $\mu$sec & \% & $\mu$sec & \% \\
        \hline
        \ce{ZnWO4}         & 8.2   & 11 & 27.9 & 89 & 
        \ce{ZnWCdMoO4(Eu)} & 9.3   & 15 & 28.5 & 85 \\ 
        \ce{ZnWPbMoO4(Eu)} & 12.5  & 24 & 32.0 & 76 &
        \ce{ZnWCdMoO4(Ce)} & 8.5   & 13 & 28.3 & 87 \\
        \ce{ZnWPbMoO4(Sm)} & 12.6  & 23 & 32.0 & 77 &
        \ce{ZnWCdMoO4(Sc)} & 8.7   & 14 & 28.3 & 86 \\
        \ce{ZnWPbMoO4(Pr)} & 13.1  & 26 & 32.5 & 74 &
        \ce{ZnWCdMoO4(Yt)} & 8.4   & 11 & 28.1 & 89 \\ 
        \ce{ZnWCdMoO4}     & 10.9  & 15 & 29.3 & 85 &
        \ce{ZnWCdMoO4(Nb)} & 11.3  & 16 & 29.0 & 84 \\
        \hline
    \end{tabular}
    \label{tab:decay}
\end{table} 
All the crystals show decay characteristics (at least, for $662$ keV photons) close to the values obtained for the reference \ce{ZnWO4} crystal. Although, some tendency against higher decay time values and  intensitites for the fast components is observed for the \ce{ZnWCdMoO4} crystals and, especially, for the \ce{ZnWPbMoO4} sample. All measurements are performed at room temperature.  

\section{Light yield relative to pure \ce{ZnWO4}}
\label{sec:light}
\begin{figure}[h]
    \centering
    \begin{subfigure}{0.5\textwidth}
    \centering
    \includegraphics[width=\textwidth]{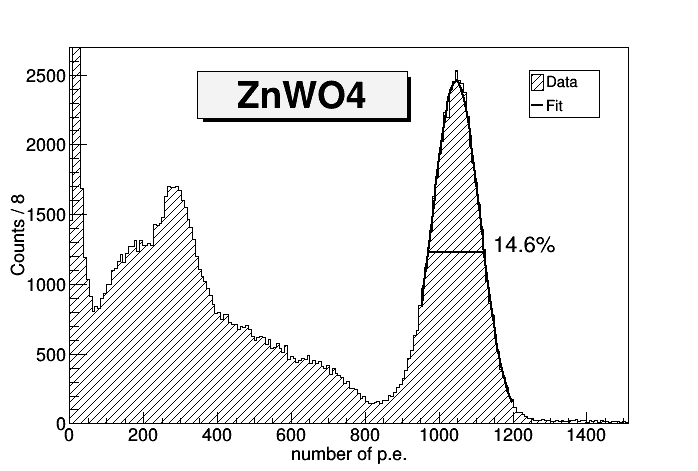}
    \caption{}
    \end{subfigure}%
    \begin{subfigure}{0.5\textwidth}
    \centering
    \includegraphics[width=\textwidth]{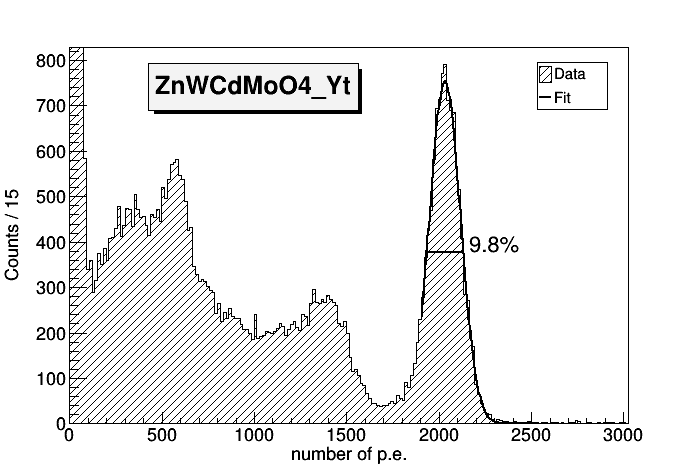}
    \caption{}
    \end{subfigure}
    \caption{${}^{137}{\rm Cs}$ energy spectra measured with (a) \ce{ZnWO} and (b) \ce{ZnCdWO(Yt)} crystals. The full absorption peak is fitted to the Gaussian function.}
    \label{fig:spectra}
\end{figure}

The light yield of the scintillation counter is its important characteristic. In this study the light yield of all crystals is measured relative to the pure \ce{ZnWO4} sample. 
Our setup with the calibrated PMT allows us to measure the number of photoelectrons corresponding to the full absorption of $662$ keV photons inside the crystal. The signal shape is integrated within $30$ $\mu$sec and integral spectrum in terms of number of photoelectrons is obtained. $^{137}{\rm Cs}$ energy spectra of the pure \ce{ZnWO4} sample and \ce{ZnWCdMoO4(Yt)} crystal with the highest photoelectron yield among other samples are shown in figure~\ref{fig:spectra}. 
Total absorption peaks are clearly seen and fitted by the Gaussian function to extract the photoelectron yield. 
The relative energy resolution of $14.6\%$ for the \ce{ZnWO4} and $9.8\%$ for the \ce{ZnWCdMoO4(Yt)} is calculated as a full width at half maximum (FWHM) of the photopeak divided by the number of photoelectrons.  
The total resolution includes the scintillator resolution and the statistical contribution related to the variation of the number of photoelectrons produced at the photocathode. 

The spectra for all studied samples are similar to that of the \ce{ZnWCdMoO4(Yt)}. Their characteristics are shown in table~\ref{tab:spectra}.
\begin{table}[h]
    \centering
    \caption{Light yield characteristics and scintillator resolutions of the \ce{ZnWPbMoO4} and \ce{ZnWCdMoO4} samples. The result for the reference \ce{ZnWO4} is also shown.}
    \begin{tabular}{|l|c|c|c|}
    \hline
\multirow{2}{*}{Crystal} & \multirow{2}{*}{$N_{\rm p.e}$/MeV} & Relative  & Scintillator \\
& & light yield & resol, \% \\  \hline
\ce{ZnWO4}   & 1580 & 1.0 & 12.2  \\
\ce{ZnWPbMoO4(Eu)} & 2705 & 1.7 & 8.1 \\
\ce{ZnWPbMoO4(Sm)} & 2535 & 1.6 & 8.5 \\
\ce{ZnWPbMoO4(Pr)} & 2491 & 1.6 & 10.6 \\
\ce{ZnWCdMoO4} & 2799 & 1.8 & 7.9 \\
\ce{ZnWCdMoO4(Eu))} & 3012 & 1.9 & 9.3 \\
\ce{ZnWCdMoO4(Ce)} & 2971 & 1.9 & 9.2 \\
\ce{ZnWCdMoO4(Sc)} & 2992 & 1.9 & 9.3 \\
\ce{ZnWCdMoO4(Yt)} & 3066 & 1.9 & 7.9 \\
\ce{ZnWCdMoO4(Nb)} & 2932 & 1.9 & 7.8 \\
\hline
    \end{tabular}
    \label{tab:spectra}
\end{table}

Light yields are defined relative to the light yield of the reference \ce{ZnWO4} crystal. In the first approximation, the spectral sensitivities of the studied samples are considered to be identical with the \ce{ZnWO4} as it is shown in section~\ref{sec:emission}. 
Light collection efficiencies of the studied samples are estimated at the level of $50-60\%$ based on the ratio of values measured with and without optical grease.  
In such a way, a ratio of light yields is estimated as a ratio of photoelectron yields (see table~\ref{tab:spectra}).

The light yields of the new mixed crystals are enhanced relative to the \ce{ZnWO4} by a factor of $1.5$ for the \ce{ZnWPbMoO4} crystals and two for the \ce{ZnWCdMoO4}. Their energy resolutions are also significantly improved from $12.2\%$ for the reference \ce{ZnWO4} to $7.8\%$ for the 
\ce{ZnWCdMo4(Nb)}. All these measurements are performed at room temperature with a ${}^{137}{\rm Cs}$ source of $662$ keV photons.

\section{Emission spectrum}
\label{sec:emission}
Emission spectrum of the studied crystals is estimated using a set of five color glass cut-on filters Newport FS-C, which are placed on the window of the PMT. The scintillator is put on the filter and irradiated by a ${}^{239}{\rm Pu}$ source of $\alpha$ particles with the energy of $5156$ keV that provides a single peak clearly distinguished even for low light signal. To avoid the $\alpha$ particle absorption in the teflon coverage, the small hole about $1$ mm${}^2$ is made in the centre of the top face.

The relative light yield with each filter $F_i$ is measured as $L_i=A_i/A_0$, where $A_i$ and $A_0$ are energy spectrum peak positions with and without filter $F_i$, respectively. Then, the relative light yield $s_{ij}$ in the wavelength range $[\lambda_i,\lambda_j]$ ($j=i+1$) is given by
\begin{equation}
s_{ij} = 
(L_i/R_i - L_j/R_j)/\int_{\lambda_i}^{\lambda_j}Q(\lambda)d\lambda,
\end{equation}  
where $R_i$ is a transmission coefficient of the filter $F_i$ at $\lambda > \lambda_i$ and
$Q(\lambda)$ is the PMT quantum efficiency.

\begin{figure}[h]
    \centering
    \begin{subfigure}{0.5\textwidth}
    \centering
    \includegraphics[width=\textwidth]{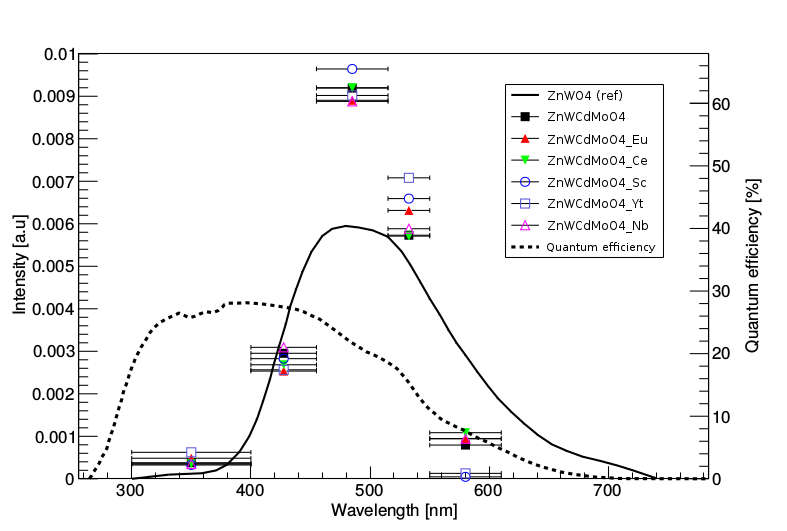}
    \caption{}
    \end{subfigure}%
    \begin{subfigure}{0.5\textwidth}
    \centering
    \includegraphics[width=\textwidth]{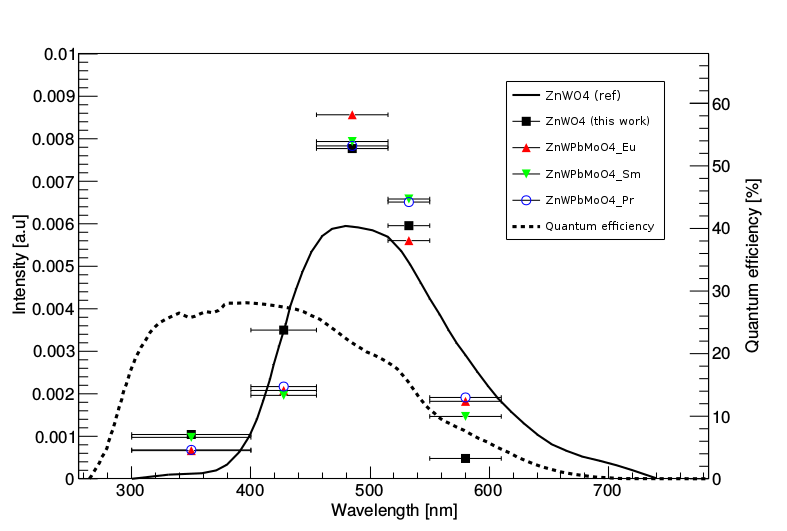}
    \caption{}
    \end{subfigure}
    \caption{Emission spectra of (a) \ce{ZnWCdMoO4} and (b) \ce{ZnWPbMoO4} samples. The spectrum of the reference \ce{ZnWO4} crystal is also measured. The solid line shows the known emission spectrum of \ce{ZnWO4} taken from ref.~\cite{mikhailik}. The dashed line corresponds to the PMT quantum efficiency curve.}
    \label{fig:emission}
\end{figure}

The resulting emission spectra are shown in figure~\ref{fig:emission}.
The emission spectrum of pure \ce{ZnWO4} crystal is taken from ref.~\cite{mikhailik}. 
We can see that all spectra peak at value lying in the range between $460$ nm and $520$ nm where the maximum intensity of the pure \ce{ZnWO4} ($480$ nm) is achieved. The PMT quantum efficiency is superimposed on the plot.

\section{Conclusion}
\label{sec:final}
Tungstate and molybdate crystals are considered as promising materials for cryogenic experiments hunting on WIMP particles or $0 \nu 2 \beta$ decays. Several new \ce{(Zn_{0.9} Pb_{0.1})(W_{0.9} Mo_{0.1})O4} and \ce{(Zn_{0.9} Cd_{0.1})(W_{0.9} Mo_{0.1})O4} mixed crystals doped with of \ce{Eu, Sm, Pr, Ce, Sc, Yt} and \ce{Nb} and produced using low-thermal-gradient Czochralski process are studied. Their scintillation characteristics are measured at room temperature. 
Light yields, energy resolutions and scintillation decay times are obtained with ${}^{137}{\rm Cs}$ source of gamma quanta while spectral sensitives are evaluated with ${}^{239}{\rm Pu}$ source of alpha particles. 

Analysis of emission spectra and decay time components shows the properties similar to the pure \ce{ZnWO4} crystal. However, the light yields of the samples are increased by a factor of up to two relative to the reference \ce{ZnWO4} scintillator (with the light yield of about $10$ ph./keV), especially for \ce{(Zn_{0.9} Cd_{0.1})(W_{0.9} Mo_{0.1})O4} samples. Their energy resolutions are also significantly improved up to $7.8\%$.
Other studies of their performance are required at low temperatures to understand a benefit of their potential application in cryogenic experiments. 

\acknowledgments
The work of E.N.G and V.A.M is supported by the Ministry of Education and Science of the Russian Federation (grant No.~3.3726.2017).
The work of D.V.M is supported by the Russian Federation Government (grant No.~14.W03.31.0026).


\end{document}